\documentstyle[12pt,epsf]{article}

\newcommand{\bc}{\begin{center}}
\newcommand{\ec}{\end{center}}
\newcommand{\bd}{\begin{displaymath}}
\newcommand{\ed}{\end{displaymath}}
\newcommand{\be}{\begin{equation}}
\newcommand{\ee}{\end{equation}}
\newcommand{\ba}{\begin{array}}
\newcommand{\ea}{\end{array}}
\newcommand{\bt}{\begin{tabular}}
\newcommand{\et}{\end{tabular}}

\newcommand{\ov}{\overline}
\newcommand{\bp}{\begin{picture}}
\newcommand{\ep}{\end{picture}}
\newcommand{\bfi}{\begin{figure}}
\newcommand{\efi}{\end{figure}}

\def\fun#1#2{\lower3.6pt\vbox{\baselineskip0pt\lineskip.9pt
\ialign{$\mathsurround=0pt#1\hfil##\hfil$\crcr#2\crcr\sim\crcr}}}

\newcommand{\tg}{\mbox{tg }}
\newcommand{\vp}{\varphi}
\newcommand{\vt}{\vartheta}
\newcommand{\ds}{\displaystyle}
\newcommand{\ve}{\varepsilon}

\begin{document}

\title{\bf Mass and CKM Matrices of Quarks and Leptons,
            the Leptonic CP-phase in Neutrino Oscillations }

\date{ITEP, Moscow, 117259 B.Cheremushkinskaya 25}
\author{
D.A.Ryzhikh\thanks{ryzhikh@vitep5.itep.ru} and
K.A.Ter--Martirosyan\thanks{termarti@heron.itep.ru}.} \maketitle

\begin{abstract}
A general approach for construction of quark and lepton mass matrices
is formulated. The hierarchy of quarks and charged leptons ("electrons")
is large, it leads using the experimental values of
mixing angles to the hierarchical mass matrix slightly deviating from one's
suggested earlier by Stech and including naturally the CP-phase.

The same method based on the rotation of generation numbers in the diagonal
mass matrix is used in the electron-neutrino sector of theory, where neutrino
mass matrix is determined by the Majorano see-saw approach. The hierarchy of
neutrino masses, much smaller than for quarks, was used including all
existing (even preliminary) experimental data on neutrinos mixing.

The leptonic mass matrix found in this way includes not known value of the
leptonic CP-phase. It leads to a large $\nu_\mu\nu_\tau$ oscillations and
suppresses the $\nu_e\nu_\tau$ and also $\nu_e \nu_\mu$ oscillations. The
explicit expressions for the probabilities of neutrino oscillation were obtained
in order to specify the role of leptonic CP-phase. The value of time
reversal effect (proportional to $\sin\delta'$) was found to be small
$\sim 1\%$. However, a dependence of the values of
$\nu_e \nu_\mu, \nu_e \nu_\tau$ transition probabilities, averaged over
oscillations, on the leptonic CP-phase has found to be not small - of order of
tens percent.
\end{abstract}

\thispagestyle{empty}

\newpage

\pagenumbering{arabic}

{\bf 1.} {\bf Introduction}\\

Serious efforts have been invested recently in the natural understanding of
experimental results on neutrino oscillations. They have shown that
neutrinos of three generations have, perhaps, non vanishing small masses.
The heaviest of them, the neutrino of the third generation, seems to have
 a mass of the order of $\sim (1/20)$eV and, as the Super Kamiokande data on
atmospheric neutrinos show, has the maximal possible mixing with the neutrino of
 the second generation, and may be, also not too small mixing with that
of the first generation. This was not expected a priori, since all similar
 mixing angles of quarks are small.

It is the challenge of modern particle physics to include naturally these
results into the framework of Grand Unification Theory together with the data
 on quark masses and mixing angles. For the quarks these angles are small and
are known already [1].

We begin this paper by reminding the well-known picture of masses,
CP-phase and mixings for the quark sector of a theory. A general method will be
developed which allows one to construct consistently the $3\times 3$
mass-matrix and the CKM mixing matrix for quarks. The same general approach will
be used later for the electron-neutrino sector of a theory.

Let us consider, as a useful introduction to a consistent theory of
quark and lepton masses and mixings, a simple phenomenological approach
suggested by B.~Stech [2]. He has noticed the following quark and charged
lepton ("electrons") mass hierarchies:
\be
m_t:m_c:m_u\simeq 1:\sigma^2:\sigma^4~,~~~
m_b:m_s:m_d\simeq 1:\frac{1}{2}\sigma:8\sigma^3~,~~~
m_\tau:m_\mu:m_e\simeq 1:\sigma:\frac{3}{2}\sigma^3,
\ee
with a very small $\sigma^2 \simeq 1/300,\: \sigma \simeq 0.058$. He has also
introduced the following mass matrices which reproduced approximately
the masses of all the quarks as well as theirs mixing angles:
\be
\hat M_u=
\left(\ba{ccc}
0 & \frac{1}{\sqrt{2}}\sigma^3\eta & \sigma^2\eta \\
\frac{1}{\sqrt{2}}\ds\sigma^3\eta^+ &-\frac{1}{2}\sigma^2
& \frac{1}{\sqrt{2}}\sigma \\
\sigma^2\eta^+ &\frac{1}{\sqrt{2}}\sigma & 1 \ea\right)m_t,\quad
\hat M_d=
\left(\ba{ccc} 0 & a_d\sigma^3 & 0\\
a_d\sigma^3 & -\frac{\ds\sigma^2}{\ds 2} & 0\\
0 & 0 & \sigma
\ea\right)\frac{m_b}{\sigma}
\ee
\be
\hat M_e= \left(\ba{ccc}
0&a_e\sigma^3&0\\
a_e\sigma^3&-\sigma^2&0\\
0&0&\sigma\ea \right)\frac{m_{\tau}}{\sigma}
\ee

Here $\eta=e^{i\delta}$ represents the CP violating phase $\delta$ in the quark
sector, while the values of the constants
$a_d\simeq 2,$ $a_e\simeq \sqrt\frac{\ds 3}{\ds 2}$ correspond to the best fit
of all masses of quarks and electrons (their central values, see below).

The diagonalization of the matrices (2) and (3) by means of an unitary matrix
$\hat U_a$:
\be
\hat M_a^{diag}=\hat U_a \hat M_a \hat U^+_a \qquad a=u,d,e,
\ee
(where $\hat U_a\hat U_a^+=\hat U_a^+\hat U_a=1$), reproduces simultaneously
both the experimental (central) values of running masses of quarks and electrons
and all quark mixing angles (their sines):
\mbox{$s_{12}=\sin\vartheta^q_{12},\; s_{23}=\sin\vartheta^q_{23},\;
s_{13}=\sin\vartheta^q_{13}$} in the CKM matrix:
\be
\hat V_{CKM}=\hat U^+_u\hat U_d=
\left( \ba{ccl}
c_{12}c_{13}&s_{12}c_{13}&s_{13}e^{-i\delta'}\\
-s_{12}c_{13}-c_{12}s_{23}s_{13}e^{i\delta'}&c_{12}c_{23}-
s_{12}s_{23}s_{13}e^{i\delta'}&s_{23}c_{13}\\
s_{12}s_{23}-c_{12}c_{23}s_{13}e^{i\delta'}&-c_{12}s_{23}-
s_{12}c_{23}s_{13}e^{i\delta'}&c_{23}c_{13}\\
\ea
\right).
\ee
Here, according to experimental data [1,6], one has
\footnote[1]{The upper index $q$ is used to emphasize that the angles
$\vt_{ij}=\vt^q_{ij}$, or $s_{ij}=\sin\vt^q_{ij}$ determine
just by the quarks and not by the leptonic mixings. The corresponding
leptonic angles and their sines are denoted below by $\vt^l_{ij}$ and by
$s_{ij}=\sin\vt^l_{ij}$ without any upper index (see sect. $\bf 3$
and $\bf 4$). The leptonic CP-phase is denoted (in sect.$\bf {3,4}$)
by $\delta'$.}:
\be
\ba{c}
s_{12}\simeq\sin\vartheta^q_{12}\simeq 0.221\pm
0.004,\;
s_{23}=\sin\vartheta^q_{23}=0.039\pm 0.003,\\
s_{13}=\sin\vartheta^q_{13}\simeq s_{12}s_{23}|R|=0.0032\pm 0.0014,\;
|R|=s_{13}/s_{12}s_{23}\simeq 0.38\pm 0.19\\
\mbox{ and }\delta=\pi/2\pm\pi/4. \hspace{8.2cm}
\ea
\ee
The factors $1/\sqrt{2}$, $1/2$, $2$, $\sqrt{3/2}$, etc. in Eq.(2),(3) were
adjusted so as to reproduce all these mixing angles and all the
masses of  quarks and leptons. As has been mentioned, for quarks the mixing
angles are small, i.e. $s_{13}\ll s_{23}\ll s_{12}\ll 1$ and therefore the
cosines $c_{13},c_{23},c_{12}$  of them can be put approximately equal
to unity.

The major ideas of Stech's approach were: 1) to consider the running masses
of all quarks at the same scale (instead of the pole ones) and 2) to obtain
these masses and mixing angles in terms of few parameters. However, the
matrices (2) and (3) have been written phenomenologically, just by hand.

Below we develop an approach to quark and lepton masses and mixings
through the following steps:

a) A general method is suggested which allows one to construct the quark mass
matrices consistently (without using Eq.(2)) in terms of their masses and
mixing angles (6).  The resulting mass matrices coincide with those given by
Eq.(2). They are diagonalized in an analytical form of decomposition in
powers of small parameter $\sigma$ and the quark CKM matrix is reproduced
analytically.

b) The same approach is applied to the reconstruction the neutrino mass
matrices $\hat M_{\nu}$. This is performed using the Eq.(4) and the unitary
matrix $\hat U_{\nu}$. Preliminary results on masses of neutrinos and their
mixing angles, extracted from neutrino oscillation experiments have been used.

c) The leptonic CKM matrix $\hat V_{CKM}^l=\hat U^+_{\nu}\hat U_e$ containing
the leptonic CP-phase $\delta'$ is introduced. It leads to $\nu_{\mu}\nu_{\tau}$
oscillation of the type observed at Super-Kamiokande and also to some suppression
of $\nu_e\nu_{\mu}$ oscillation.

d) The expressions for the three type neutrino oscillations are specified
and some realistic numerical examples of oscillation dependence on the leptonic
CP-phase are presented.

Finishing this section let us note that the electron and quark masses emerge from
the usual Yukawa interaction:
$$
L'_{\mbox{\footnotesize Yuk}}=\vp_2\ov u_i h_u^{ij} q_j +
\vp_1\ov d_i h_d^{ij} q_j + \vp_1 \ov e_i h_e^{ij} l_j \; ,
$$
where our mass matrices (2) and (3) are:
$
(\hat M_u)=\langle\vp_2\rangle\hat h_u,
\quad (\hat M_d)=\langle\vp_1\rangle\hat h_d,\quad
(\hat M_e)=\langle\vp_1\rangle\hat h_e
$
and $\langle\vp_1\rangle =v\cos\beta,\:\langle\vp_2\rangle =v\sin\beta$ are
V.E.V of SUSY two neutral CP-even Higgs fields ($v=174$GeV, and $\tg\beta$ are
two well known SUSY parameters) and $h_e^{ij}\sim h_d^{ij}\sim h_u^{ij}\sim 1$
are the Yukawa coupling constants.

The scale of neutrino masses
($10^{11}-10^{12}$ times smaller than the electron and quark masses)
can be a result of the see--saw approach [3--5]. It leads after the
integrating out of the corresponding heavy states (e.g. the right handed
Majorano neutrinos with heavy masses of an order of $M\sim 10^{14}-10^{15}$GeV)
to the appearance of higher order effective Majorano mass operator (see
[3],[5]):
$$
(\nu_i(\hat M_{\nu})^{ij}\nu_j)=(\vp_2^2/M)(\nu_ih^{ij}\nu_j),
$$
where $h^{ij}\sim 1$ are the neutrino coupling constants and
$\hat M_\nu=(\vp_2^2/M)\hat h$ is the neutrino mass matrix.\\

{\bf 2. Quarks mass matrices and the analytical form of the CKM matrix.}\\

The running of u- and d-quark masses calculated [7] in the first (dashed
curves) and in the fourth (solid curves) order of QCD perturbation
theory are shown [8] in Fig.~1.
The quark running masses $m_i(\mu)$ are related in $\overline{MS}$
renormalization scheme at the scale $\mu=M_i$ to their pole masses $M_i$ by the
well known relation:
$$
m_i(M_i)={M_i}/[1+\frac{4\alpha_s(M_i)}{3\pi}+
K(\frac{\alpha_s(M_i)}{\pi})^2],
$$
where K=11,2.
The scale $\mu=M_t=174.4$ GeV is the most natural for the SUSY Standard
Model. The running quark masses at this scale, (see in Fig.~1) have the
following values (in GeV's):
\be
\ba{lll}
m_u(M_t)=(0.21\pm0.1)\cdot 10^{-2},&
m_d(M_t)=(0.42\pm 0.21)\cdot 10^{-2},&
m_e=0.511\cdot 10^{-3},\\
m_c(M_t)=0.59\pm 0.07,&
m_s(M_t)=0.082\pm 0.041,&
m_\mu=105.66\cdot 10^{-3},\\
m_t(M_t)=163\pm4,&
m_b(M_t)=2.80\pm 0.40,&
m_\tau=1.777 \pm 0.0003.\ea
\ee
These values differ strongly from the ones used in Ref.[2] which are
determined at a very small scale $\mu=1$ GeV. However, the Stech's
relation between them still holds.

Note that the matrices $\hat M_d$ and $\hat M_e$ in Eqs.(2),(3)
have the block structure and their diagonalization is trivial.
In fact, for any $2\times2$ matrix
$
\hat m=\left(
\raisebox{10pt}{\makebox[14pt][c]{$a$}}
\raisebox{-10pt}{\makebox[0pt][r]{$\rho\;$}}
\raisebox{10pt}{\makebox[14pt][r]{$\rho\,$}}
\raisebox{-10pt}{\makebox[0pt][r]{$b\,$}}
\right),
$
which $\hat M_d$ and $\hat M_u$ contain in their left-upper part, one has:
\be
\hat m_{diag}=\hat u \hat m \hat u^+=
\left(
\raisebox{10pt}{\makebox[11pt][r]{$\mu_1$}}
\raisebox{-10pt}{\makebox[0pt][r]{$0\;\:$}}
\raisebox{10pt}{\makebox[18pt][r]{$0\;$}}
\raisebox{-10pt}{\makebox[0pt][r]{$\qquad\mu_2$}}
\right),
~~\hat u=\left(
\raisebox{10pt}{\makebox[14pt][r]{$c$}}
\raisebox{-10pt}{\makebox[0pt][r]{$-s$}}
\raisebox{10pt}{\makebox[20pt][r]{$s\,$}}
\raisebox{-10pt}{\makebox[0pt][r]{$c\,$}}
\right),
~~\mu_{1,2}=\frac{1}{2}(a+b\pm \sqrt {(a-b)^2+4\rho^2}\; ).
\ee
Here
$t_{2\vartheta}=\tan 2\vartheta=2\rho/(a-b)$ and
$s=\sin \vartheta=\frac{1}{\sqrt 2}
(1-(1+t^2_{2\vartheta})^{-\frac{1}{2}})^{\frac{1}{2}}$,
$c=\cos\vartheta =$ \mbox{$\frac{1}{\sqrt 2}
(1+(1+t^2_{2\vartheta})^{-\frac{1}{2}})^{\frac{1}{2}}$}, and $\vt$ is the
mixing angle.
In Eqs.(2),(3) for the matrices $\hat M_d$ and $\hat M_e$ one has $a=0$ and the
mixing angles $\vartheta_d, \vartheta_e$ are small since
$t_{2\vartheta_d}=2\frac{\ds 2\sigma^3}{\ds\sigma^2/2}=8\sigma<1$
and
$t_{2\vartheta_e}=2\ds{\sqrt{\frac{3}{2}}\frac{\sigma^3}{\sigma^2}}=
\sqrt 6\sigma$ is even smaller.
Due to the block structure of $\hat M_d$ and $\hat M_e$
the unitary matrices $\hat U_d$ and $\hat U_e$
also have the following block structure:
\be
\hat
U_d=\left(\ba{clr}c_d&s_d&0\\-s_d&c_d&0\\0&0&1\ea\right),
\quad
\hat
U_e=\left(\ba{rcc}c_e&s_e&0\\-s_e&c_e&0\\0&0&1\ea\right)
\ee
where $s_d=\frac{\ds 1}{\ds\sqrt 2}
(1-(1+(8\sigma)^2)^{-\frac{1}{2}})^\frac{1}{2}\simeq 4\sigma(1-24\sigma^2)
\simeq 0.214\, ,\quad
s_e=\frac{\ds 1}{\ds\sqrt 2}(1-(1+6\sigma^2)^{-\frac{1}{2}})^\frac{1}{2}
\simeq \sqrt\frac{\ds 3}{\ds 2}\sigma(1-(\frac{\ds 3}{\ds
2}\sigma)^2)\simeq 0.0705$ up to the terms of the order of $\sigma^4$.

Let us note that Stech's  matrices $\hat M_d,\hat M_e$ in Eqs.(2),(3) can be
reconstructed using their diagonal form (i.e. the physical masses of d-quarks
and electrons):
\be
\hat M^{diag}_d=
\left(
\raisebox{10pt}{\makebox[12pt][l]{$m_d$}}
\raisebox{0pt}{\makebox[17pt][c]{$-m_s$}}
\raisebox{-10pt}{\makebox[17pt][r]{$m_b$}}
\right)\, ,\quad
\hat M^{diag}_e=
\left(
\raisebox{10pt}{\makebox[12pt][l]{$m_e$}}
\raisebox{0pt}{\makebox[17pt][c]{$-m_\mu$}}
\raisebox{-10pt}{\makebox[17pt][r]{$m_\tau$}}
\right),
\ee
by means of Eq.(4), which states:
\be
\hat M_d=\hat U^+_d\hat M_d^{diag}\hat U_d, \qquad
\hat M_e=\hat U^+_e\hat M_e^{diag}\hat U_e.
\ee
The matrices $\hat U_d=\hat O_{12}^d$ (or $\hat U_e=\hat O^e_{12}$) in Eq.(9)
can be considered as rotating the 12 generations of d-quarks (or electrons).

A bit more complicated (but much more instructive) is the diagonalization of
the matrix~$\hat M_u$. Similarly to $\hat M_d$ and $\hat M_e$ this matrix can be
represented as :
\be
\hat M_u=\hat U^+_u\hat M_u^{diag}\hat U_u,~~~
\mbox{ where } \hat U_u^+=\hat O^+_{23}\hat O^+_{13}(\delta)(\hat O_{12}^u)^+
\ee
Here the matrices:
\be
(\hat O_{12}^u)^+=
\left( \ba
{rrc}c_u&-s_u&0\\s_u&c_u& 0\\0&0&1
\ea \right),~~
\hat O_{13}^+(\delta)=
\left( \ba {ccc}
c_{13}&0&s_{13} e^{-i\delta}\\0&1&0\\-s_{13} e^{i\delta} &0&c_{13}
\ea \right),
~~\hat O_{23}^+=
\left( \ba
{lrc}1&0&0\\0&c_{23}&s_{23}\\0&-s_{23}&c_{23}
\ea \right)
\ee
rotates 12, 13 and 23 generations, respectively. The 13 rotation
includes naturally the complex phase $\delta$ which violates the CP-parity
conservation of a theory. It can not be removed by a trivial phase
transformation of the u-quark or t-quark  fields. However, the value of $s_u$
turns out to be very small $(s_u\sim\sigma^5 \ll 1)$ and one can really put
$\hat O^u_{12}\simeq \hat 1$. The quark CKM matrix is:
\be
\hat V_{CKM}^q=
\hat U^+_u\hat U_d=\hat O^+_{23}\hat O^+_{13}(\delta) \hat O^+_{12},
~~\mbox{ where }
\hat O^+_{12}=(\hat O^u_{12})^+\hat O^d_{12}\simeq \hat O_{12}^d.
\ee
Here in general
$
(\hat O_{12}^u)^+= \left(
\ba{rcc} c_{12}&s_{12}&0\\
-s_{12}&c_{12}& 0\\
0&0&1
\ea \right),~~
$
with $s_{12}=s_{12}^dc_u-c_{12}^ds_u=\sin(\vt_d-\vt^u_{12})\simeq\sin\vt_d$
since $s_u\sim\sigma^5$ (see below) is neglegibly small.

Thus for the given upper quark masses
$
\hat M^{diag}_u=
\left(
\raisebox{9pt}{\makebox[14pt][c]{$m_u$}}
\raisebox{0pt}{\makebox[14pt][c]{$m_c$}}
\raisebox{-9pt}{\makebox[14pt][c]{$m_t$}}
\right)
$
and the given quark mixing angles (from the quark CKM mixing matrix
$\hat V_{CKM}^q$, see below) the form of $\hat U^+_u$ is fixed:
\be
\hat U^+_u =\hat O^+_{23}\hat O^+_{13}(\delta)=
\left(
\ba{ccc}
c_{13} & 0 & s_{13}e^{-i\delta}\\
s_{13}s_{23}e^{i\delta} & c_{23} & c_{13}s_{23}\\
-s_{13}c_{23}e^{i\delta} & -s_{23} & c_{13}c_{23}\\
\ea\right)
\ee
The same matrix $\hat U^+_u$ can be obtained in the form of decomposition in
powers of $\sigma$ (up to $\sigma^4$ terms) by a direct calculation:
\be \hat U^+_u =\left(
\ba{ccc}
1-\frac{\ds\sigma^4}{\ds 2} & 0 &
\sigma^2(1-\frac{\ds\sigma^2}{\ds4})e^{-i\delta}\\
-\frac{\ds\sigma^3}{\ds\sqrt 2}e^{i\delta} & 1-\frac{\ds\sigma^2}{\ds 4} &
\frac{\ds\sigma}{\ds\sqrt
2}(1-\frac{\ds 5}{\ds 4}\sigma^2)\\
-\sigma^2(1-\frac{\ds\sigma^2}{\ds 2})e^{i\delta} &
-\frac{\ds\sigma}{\ds\sqrt 2}(1-\frac{\ds 5}{\ds 4}\sigma^2) &
1-\frac{\ds\sigma^2}{\ds 4}\\
\ea \right)
\ee
Multiplying $\hat U^+_u$ in this form by $\hat U_d$ one obtains with the same
accuracy the following CKM quark matrix $\hat V_{CKM}^q=\hat U_u^+\hat U_d$:
\be
\hat V_{CKM}=\left(
\ba{ccc}
(1-\frac{\ds\sigma^4}{\ds
2})c_{12}& (1-\frac{\ds\sigma^4}{\ds 2})s_{12} &
\sigma^2(1-\frac{\ds\sigma^2}{\ds 4})e^{-i\delta}\\
-(1-\frac{\ds\sigma^2}{\ds 4})s_{12}-\frac{\ds\sigma^3}{\ds\sqrt 2}c_{12}
e^{i\delta}&
(1-\frac{\ds\sigma^2}{\ds 4})c_{12}-\frac{\ds\sigma^3}{\ds\sqrt 2}s_{12}
e^{i\delta} &
\frac{\ds\sigma}{\ds\sqrt 2}(1-\frac{\ds 5}{\ds 4}\sigma^2)\\
\frac{\ds\sigma}{\ds\sqrt 2}(1-\frac{\ds5}{\ds 4}\sigma^2)s_{12}
-\sigma^2c_{12}e^{i\delta}&
-\frac{\ds\sigma}{\ds\sqrt 2}(1-\frac{\ds 5}{\ds 4}\sigma^2)c_{12}
-\sigma^2s_{12}e^{i\delta}&
1-\frac{\ds \sigma^2}{\ds 4}
\ea \right)
\ee
Comparing  the CKM matrix (5) with Eq.(17) one finds:
\be
\ba{c}
s_{12}=s_d\simeq4\sigma(1-24\sigma^2)\approx 0.215,\quad
s_{23}=\frac{\ds \sigma}{\ds \sqrt 2}(1-\frac{\ds 5}{\ds 4}\sigma^2)
\approx 0.0408,\\
s_{13}=s_{12}s_{23}|R|=\sigma^2(1-\sigma^2/4)\approx 3.36\cdot 10^{-3}
\ea
\ee
implying that $|R|=\frac{\ds s_{13}}{\ds s_{12}s_{23}}=
\frac{\ds1} {\ds\sqrt 8(1-25\sigma^2\ds)}\ds\approx 0.38$

Calculating $\hat M_u$ by means of Eq.(12) one first obtains
$\hat M_u'=\hat O_{13}^+M_u^{diag}\hat O_{13}$ and then finds
$\hat M_u=\hat O_{23}^+\hat M'_u\hat O_{23}$ in the form:
\be
\hat M_u=\left(
\ba{ccc}
0&s_{13}s_{23}e^{-i\delta}&s_{13}e^{-i\delta}\\
s_{13}s_{23}e^{i\delta}&s_{23}^2+\frac{\ds m_2}{\ds m_3}& s_{23}\\
s_{13}e^{i\delta}&s_{23}&1\\
\ea
\right) m_3
\ee
up to the terms of an order of $\sigma^4$

Eq.(19) really reproduces the Stech's matrix $\hat M_u$ in Eq.(2) only for
negative sign of $m_2=-|m_2|$; for positive $m_2=|m_2|$ the ratio
$\mu_{22}=(\hat M_u)_{22}/m_t$ is $\frac{\ds 3}{\ds 2}\sigma^2$ instead of
$-\frac{\ds\sigma^2}{\ds 2}$ in Eq.(2). Considering higher order
$\sigma^2$ corrections to the matrix (19) one finds that
$\mu_{11}=(\hat M_u)_{11}/m_t$ turns out to be $2\sigma^4$ for positive
$m_u=|m_u|$ and is much smaller, $-\frac{\ds\sigma^6}{\ds 2}$, for negative
lightest eigenvalue when $m_u=-|m_u|$.

Therefore, literally the Stech's matrix $\hat M_u$ is reproduced in the form (2)
only for negative $m_1$  and $m_2$, e.g. at:
$
\hat M^{diag}_u=
\left(
\raisebox{12pt}{\makebox[18pt][l]{$-|m_u|$}}
\raisebox{0pt}{\makebox[18pt][l]{$\,-|m_c|$}}
\raisebox{-11pt}{\makebox[30pt][r]{$m_t$}}
\right)
$

Thus, the u-quark mass matrix $\hat M_u$ can be reproduced by the former of
Eqs.(12) using  the $\hat U_u$ and  $\hat U_u^+$ matrices in the form of the
 latter of Eqs.(12) (or by Eq.(15) if $(\hat O^u_{12})^+\simeq 1$). Actually
this method will be very useful later for the restoration of the neutrino mass
matrix.

All these results are in a good agreement with the experimental data presented
above.Also the diagonalisation (4) of $\hat M_u\mbox{ and }\hat M_d$ with
the help of $\hat U_u\mbox{ and }\hat U_d$ matrices leads to the following
reasonable quark masses (in GeV's, as in Eq.(1)):
\be
\begin{array}{ll}
\left\{\begin{array}{l}
m_u=\sigma^4(1-\sigma^4)m_t(M_t)\approx 2.00\cdot 10^{-3}\\
m_c=\sigma(1-\sigma^2/2)m_t(M_t)\approx 0.56\\
m_t=(1-\sigma^2/2)m_t(M_t)\approx m_t(M_t)=163
\end{array}\right. &
\left\{\begin{array}{l}
m_d=\frac{\ds\sigma}{\ds 4}(\sqrt{1+(8\sigma)^2}-1)m_b=0.416\cdot 10^{-3}\\
m_s=\frac{\ds\sigma}{\ds4}(\sqrt{1+(8\sigma)^2}+1)m_b=0.085\\
m_b=m_b(M_b)=2.83
\end{array}\right.
\end{array}
\ee
These results correspond fairly well, as Stech has remarked [2], to the
experimental data (7).

The minus sign of some eigenvalues in Eq.(10) and in $\hat M_u^{diag}$ can be
easily removed by redefinition of the corresponding quark field:
$q_k \to\gamma_5 q_k$.\\ \\

{\bf 3.} {\bf Neutrino mass matrix, the leptonic CKM matrix and CP-phase}\\

To proceed further let us introduce the neutrino mass matrix
$\hat M_{\nu}$ with
a structure similar to that of $\hat M_u$. The Super Kamiokande data [9] suggest
a large $\nu_{\mu}\nu_{\tau}$ neutrino mixing i.e. large
$t_{23}=\tg 2\vt_{23}\gg 1$, or $\sin^2 2\vt_{23}=(1+t^{-2}_{23})^{-1}\simeq 1$
(i.e. $s_{23} \simeq c_{23} \simeq \frac{\ds 1}{\ds \sqrt 2}\simeq 0.71$) and
not too small value  of
$\Delta m^2_{32}=m^2_3-m^2_2 \simeq (0.59 \pm 0.20)^2 10^{-2}$eV.
As $m_3 \gg m_2$ (see below) one has:
\be
m_3 \simeq\sqrt {\Delta m^2_{32}} \simeq (0.059 \pm 0.020) \mbox{eV}
\ee

Simultaneously the atmosphere and solar neutrino data [10 - 12] (see the
discussion in Refs.[13,14]) show a large suppression of $\nu_e \nu_\mu$
and also $\nu_e \nu_{\tau}$ oscillations which have not yet been observed.
This can be a result of small mixing angles
$s_{12} \le s_{13} \ll s_{23} \sim 1/\sqrt 2$
(see Ref. [10] and also [12 - 14]):
\be
s_{12} \simeq 0.035 \pm 0.020, ~~~~s_{13}\simeq 0.25 \pm 0.10,
~~~~s_{23}=0.7 \simeq 1/\sqrt2
\ee
and some (obviously not too large) hierarchy of three neutrino masses
$m_1 \ll m_2 \ll m_3$ of a type considered above for the quarks and elections:
\be
m_3:m_2: m_1 = 1: \sigma^2_{\nu}:\sigma^4_{\nu}.
\ee
Here $\sigma^2_{\nu}$ is an unknown parameter which we can choose to be equal to
$\sigma=0.058$ to avoid the introduction of additional new parameter
$\sigma_{\nu}=\sqrt {0.058}\simeq 0.24$.  This gives $m_2 \simeq 0.34\cdot
10^{-2}$eV and together with Eq.(22) seems to be in approximate agreement
with the atmospheric neutrino data [12].

Essentially the other possibility has been suggested in a recent
paper [15], where the neutrinos $\nu_1,\nu_2\mbox{ and }\nu_3$ have been
considered with almost equal masses $m_3 \simeq m_2 \simeq m_1 \simeq 3$eV but
a large hierarchy has been introduced in their mixing angles \be s_{13} \simeq
0 \ll s_{12} \simeq s_{23} \simeq 1/\sqrt 2 \ee

This situation is not considered below as it seems more natural that
neutrino have small mass hierarchy of the type (1) like the u-quarks, but
 with much smaller power of hierarchy: $\sigma_{\nu}  \simeq 0.24\gg \sigma $.
It is very suitable to describe the situation in neutrino physics
since the condition (23) $m_2 < m_3$ (and $m_1 < m_2$) together with Eqs.(22)
suppress strongly the non observed (at least for a moment) $\nu_e \nu_{\mu}$
oscillations from electron neutrino sources on the Earth.

Therefore, let us take approximately the central values of the data given
in Eqs.(21)--(23) as a basis of our approach (see e.g. Ref.[5])
putting also $\sigma_{\nu} \simeq \sqrt{0.058}$ in  Eq.(23) and
taking the following values for the neutrino masses:
\be
M^{diag}_{\nu}=\left(
\raisebox{10pt}{\makebox[14pt][l]{$m_1$}}
\raisebox{0pt}{\makebox[14pt][c]{$m_2$}}
\raisebox{-10pt}{\makebox[14pt][r]{$m_3$}}
\right) =
\left(
\raisebox{11pt}{\makebox[20pt][l]{$0.019$}}
\raisebox{0pt}{\makebox[22pt][r]{$0.34$}}
\raisebox{-11pt}{\makebox[20pt][l]{$5.90$}}
\right)10^{-2}eV
\ee
where $m_2 \simeq \sigma^2_{\nu}m_3,\:  m_1 \simeq \sigma^2_{\nu}m_2$ and the
central value of $m_3$ has been taken from Eq.(21).

The main point of this paper is that approach of the type of Eqs.(12)--(16)
together with the electron matrix $\hat M_e$ (determined by Eq.(3)) allows one
to construct in general analytical form the neutrino mass matrix $\hat M_{\nu}$
and also the leptonic CKM matrix. The matrix $\hat M_{\nu}$ so
obtained will have all desired properties reproducing naturally Eqs.(21) -
(23) and will naturally include the CP-phase.

To this end let us remind that $\hat M_e$ in Eq.(3) has the form:
\be
\hat M_e=\hat U_e^+ \hat M_e^{diag}\hat U_e,
\ee
where
$\hat U_e=\hat O^e_{12}=
\left( \ba {rll}c_e&s_e&0\\-s_e&c_e&0\\0&0&1 \ea \right)$
has been determined in Eq.(9) with $s_e\simeq 0.0705$.
Here
$
\hat M^{diag}_e=\left(
\raisebox{10pt}{\makebox[13pt][c]{$m_e$}}
\raisebox{0pt}{\makebox[13pt][c]{$m_{\mu}$}}
\raisebox{-10pt}{\makebox[13pt][c]{$m_{\tau}$}}
\right)=
\left(
\raisebox{11pt}{\makebox[30pt][l]{$0.5175$}}
\raisebox{0pt}{\makebox[21pt][c]{$103.6$}}
\raisebox{-11pt}{\makebox[22pt][l]{$1777$}}
\right)$MeV
represents exactly the three electron
masses\footnote[1]{Let us mention that the matrix $\hat M_e$  can be slightly
modified:
$
\hat M_e\to\hat M^{\prime}_e= \left(\ba{ccc}
\lambda\sigma^4&\sqrt{\frac{\ds 3}{\ds 2}}\sigma^3&0\\
\sqrt{\frac{\ds 3}{\ds 2}}\sigma^3&-\beta\sigma^2&0\\0&0&
\sigma
\ea\right)\frac{\ds m_{\tau}}{\ds\sigma},
$
where $\beta=1+\lambda_1$ with very small $\lambda_1=0.0202308$
and $\lambda=0.0106891$ adjusted to reproduce exactly the well known
masses of all three electrons:
$\hat M_e^{'diag}=\hat U_e\hat M_e\hat U_e^+=
\left(
\raisebox{11pt}{\makebox[32pt][l]{$0.510999$}}
\raisebox{0pt}{\makebox[32pt][l]{$105.6584$}}
\raisebox{-11pt}{\makebox[28pt][l]{$1777.0$}}
\right)$MeV.
This modification will change negligibly the leptonic CKM matrix, leading to
$s_e\simeq 0.0689$ and is not important at all as the neutrino basic parameters
are determined very roughly in Eqs.(21)--(23) (see also the Conclusion).}.

The general form of the matrix $\hat M_{\nu}$ can be written quite similar to
Eq.(12):
\be
\hat M_{\nu} = \hat U_{\nu}^+ \hat M_{\nu}^{diag}\hat U_{\nu}
~~~~\mbox{with}~~~~
\hat U_{\nu}^+=\hat O_{23}^{l+} \hat O_{13}^{l+}(\delta') \hat O_{12}^{\nu+}\: ,
\ee
where
$
\hat O_{12}^{\nu}=
\left( \ba {rll}c_{\nu}&s_{\nu}&0\\-s_{\nu}&c_{\nu}&0\\ 0&0&1
\ea \right)
$
is defined similarly to $\hat O_{12}^e$ with the substitution
$s_{\nu}=\sin \vt^{\nu}_{12}$ for $s_e$ in $\hat O_{12}^e$ defined
above (the value of $\vt^{\nu}_{12}$ will be determined later on).
The matrices:
\be
\hat O_{13}^{l^+}(\delta')=
\left( \ba {ccc}
c_{13}&0&s_{13} e^{-i\delta'}\\0&1&0\\-s_{13} e^{i\delta'} &0&c_{13}
\ea \right),
~~~ \hat O_{23}^{l^+}=
\left( \ba
{lrc}1&0&0\\0&c_{23}&s_{23}\\0&-s_{23}&c_{23}
\ea \right)
\ee
represent the rotation of 13 and 23 generations of $\hat M_{\nu}^{diag}$,
respectively, and $\delta'$ is the leptonic CP-phase which has appeared here
naturally (clearly $s_{ij},c_{ij}$ from here and later on means the sines and
cosines of the leptonic and not quark mixing angles).

To calculate $\hat M_{\nu}$ in Eq.(27) in explicit form we
note that according to Eq.(8) one has
\be
\hat O_{12}^{\nu +}
\left(
\raisebox{10pt}{\makebox[14pt][l]{$m_1$}}
\raisebox{0pt}{\makebox[14pt][c]{$m_2$}}
\raisebox{-10pt}{\makebox[14pt][r]{$m_3$}}
\right)
\hat O_{12}^{\nu}= \left( \ba {rrl}a&\rho&0\\\rho&b&0\\0&0&m_3 \ea \right)
\ee
with
$a=s^2_{\nu}m_2+c^2_{\nu}m_1 \simeq m_1,\: b=c^2_{\nu}m_2+s^2_{\nu}m_1
\simeq m_2,\: \rho=-c_{\nu}s_{\nu}(m_2-m_1) \mbox{ and } |\rho|\ll m_2$
is small since $s_\nu$ in Eq.(22) is very small (actually $s_\nu\simeq
s_{12}-s_e\simeq 0.036$ see below).  Then it is easy to calculate the matrix:
$$
\hat M'=\hat O_{13}^+(\delta')
\left( \ba{rrl}
a&\rho&0\\
\rho&b&0\\
0&0&m_3\\
\ea \right) \hat O_{13}(\delta')
$$
and further to find the following result for the neutrino mass matrix
$\hat M_{\nu}=\hat O^{l+}_{23}\hat M'\hat O^{l}_{23}$:
\be \hat M_{\nu}=
\left(
\ba{rrr}
ac_{13}^2+m_3s^2_{13}&\ c_{13}s_{13}s_{23}(m_3-a)e^{-i\delta'}
&c_{23}c_{13}s_{13}(m_3-a)e^{-i\delta'}\\
&+\rho c_{13}c_{23}&-\rho s_{23}c_{13}\\
c_{13}s_{13}s_{23}(m_3-a)e^{i\delta'}&
s^2_{23}(m_3c^2_{13}+a^2s^2_{13})&
c_{23}s_{23}(m_3c^2_{23}-b+as^2_{13})\\
+\rho c_{13}c_{23}&+c^2_{23}b+\Delta_{22}&+\Delta_{23}\\
c_{23}c_{13}s_{13}(m_3-a)e^{i\delta'}&
c_{23}s_{23}(m_3c^2_{13}-b+as^2_{13})&
c^2_{23}(m_3c_{13}^2+as^2_{13})\\
-\rho s_{23}c_{13}&+\Delta_{32}&+s^2_{23}b+\Delta_{33}
\ea \right)
\ee
The values of $\Delta_{22}=-\rho s_{13}\sin{2\vt^\nu_{23}}\cos
\delta'$, $\Delta_{23}=-\rho s_{13}(e^{-i \delta'}-2s^2_{23}\cos
\delta')=\Delta^+_{32}$ and of $\Delta_{33}=-\Delta_{22}$ are small
and can be neglected. Also since $a\ll b\ll m_3,\;s_{\nu}\simeq s_{12}$
(see below) and $s^2_{12} \ll s^2_{13} \ll 1$ are very small, one
can disregard in the matrix $\hat M_{\nu}$ all the terms containing $a,b$ and
put $c_{12} \simeq c_{13} \simeq 1$.  The value of $|\rho| \simeq s_{12}m_2$ is
also very small $|\rho| \ll s_{13}m_3$, but the terms containing it in the
matrix (30) can not be omitted as that would violate the normal complex
structure of the matrix $\hat M_{\nu}$ and of CKM matrix considered below.
Therefore omitting small terms one obtains the matrix $\hat M_{\nu}$ in the
following simple form:
\be
\hat M_{\nu}=\left( \ba {ccc} s^2_{13}m_3+a c^2_{13}&
c_{13}(s_{13}s_{23}m_3e^{-i \delta'}+\rho c_{23})&
c_{13}(s_{13}c_{23}m_3e^{-i\delta'}-\rho s_{23})\\
c_{13}(s_{13} s_{23} m_3 e^{i \delta'}+\rho c_{23})&
s^2_{23}m_3+c^2_{23}m_2&
s_{23}c^2_{23}(m_3 c^2_{13}-m_2)\\
c_{13}(s_{13}c_{23}m_3 e^{i \delta'}- \rho s_{23})&
s^2_{23}c^2_{23}(m_3c^2_{23}-m_2)&
c^2_{23}m_3+s^2_{23}m_2 \ea \right).
\ee
Here, $s_{23} \simeq c_{23} \simeq 1/\sqrt 2$ and not too small values of
$s_{13}$ are determined in Eqs.(22).

In conclusion of this section let us construct the leptonic CKM matrix:
\be
\hat V^l_{CKM}= \hat U^+_{\nu} \hat U_e= \hat O^{l+}_{23}\hat O^{l+}_{13}
(\delta')\hat O^{l}_{12},
\ee
where
$$
\hat O^{l}_{12}=\hat O^{\nu +}_{12}\hat O^e_{12}=
\left( \ba {rcl}c_{12}&s_{12}&0\\-s_{12}&c_{12}&0\\0&0&1 \ea \right)
$$
and
$s_{12}=\sin (\vt_{12}^e-\vt_{12}^{\nu})=\sin \vt^l_{12}
\simeq 0.035 \pm 0.020$
as it is determined by Eq.(22). This gives for the neutrino $\nu_1\nu_2$
mixing angle: $\vt^{\nu}_{12}=\vt^e_{12}-\vt^l_{12} \simeq 0.036 \pm 0.020$
(in radians, or $(2.1 \pm 1.1)^{\circ}$). Multiplying the matrices
$\hat O_{ij}^{l +}$ in Eq (32) one obtains $\hat V_{CKM}^l$ just in the form of
Eq.(5) with $s_{23} \simeq c_{23} \simeq 1/\sqrt 2$ and $s_{12},~s_{13}$
given by Eq.(22).

Let us emphasize that the approach of this section, in particular the basic
parameters in (21),(22), and also the relation (23), are
approximate and have a model nature: also the hierarchy parameter
$\sigma_{\nu}$ is the main one and its value $\sigma_{\nu}=0.24$ is determined
only approximately by modern  experimental data.

Numerically the neutrino mass matrix (31) is represented in our approach
by:
\be
\hat M_{\nu}=
\left( \ba {ccc}0.130&0.335 e^{- i\delta'}+0.27 \cdot 10^{-2}&
0.341 e^{- i\delta'}+0.26 \cdot 10^{-2}\\
0.335 e^{i\delta'}+0.27 \cdot 10^{-2}&0.965&0.868\\
0.341 e^ {i\delta'}-0.26 \cdot 10^{-2}&0.868&1 \ea \right)m_3/1.975
\ee
and the CKM leptonic matrix (32) is
\be
\hat V^l_{CKM}=
\left( \ba {ccc}0.968&0.0339&0.25 e^{- i\delta'}\\
-0.175 e^{i\delta'}-0.025&0.714-0.061e^{i\delta'}&0.678\\
-0.178 e^ {i\delta'}+0.025&-0.699-0.0863e^{i\delta'}&0.692\ea \right)
\ee

Unfortunately instead of predicting the neutrino mixing angles $s_{12}, ~
s_{13}, ~s_{23}$ we have used their values (22) which are badly defined by
 experiment. The hierarchy parameter $\sigma_{\nu}$ remains, as has been
mentioned above, practically free and we have put it to $\sqrt {\sigma}
\simeq 0.24$ just by hand. Both the matrices $\hat O^{\nu}_{12}$ and
$\hat O^e_{12}$ represent a simple Abelian rotation: $\hat O^{\nu}_{12}$
by the angle $\vt^{\nu}_{12},\; \hat O^{\nu +}_{12}$ by the angle
 $(-\vt^{\nu}_{12})$ and $\hat O^e_{12}$ by the angle  $\vt^e_{12}$.
Therefore the product
$\hat O^{\nu+}_{12} \hat O^e_{12} \equiv \hat O^e_{12} \hat O^{\nu +}_{12}$
leads to a rotation by the angle $\vt^l_{12}=\vt^e_{12}-\vt^{\nu}_{12}$, where
$\vartheta^e_{12}\simeq 2\vartheta^l_{12}$ (the value of $s_e$ has been given
above just after Eq.(9) and $s_{12}$ in Eq.(18)) and therefore
$\vartheta^{\nu}_{12}\simeq\vartheta^l_{12}$, or  $s_{12}\simeq s_{\nu}$.

Similarly to the case of the quark CKM matrix, the most natural value of the
leptonic CP-phase leading to the largest possible CP violation can be
$\delta'=\pi /2\mbox{ or }\eta'=e^{i\delta'}=i$. This CP-phase $\delta'$ can
manifest itself in the Pontecorvo neutrino oscillation experiments.
It is very difficult to observe it now.  Below we discuss shortly
the possibility of these observations. The exact expressions for
probabilities of neutrino oscillations are given in Appendix, since they are
very cumbersome. Some of them have been obtained earlier in a number of
papers [16~-~20].\\ \\

{\bf 4.} {\bf The leptonic CP-phase in neutrino oscillations experiments}\\

Many papers were devoted to the studies, pioneered by Bruno Pontecorvo [16], of
two and of three [17~-~20],[5] neutrino oscillations.  We consider them below
shortly in order mainly to specify the role of the leptonic CP-phase [17] in
these oscillations.

Let us express $\nu_e,\nu_{\mu}$
and $\nu_{\tau}$ fields entering the weak interaction Lagrangian in
terms of neutrino states
$\nu_1,\nu_2,\nu_3$ with definite masses
$m_1,m_2,m_3$ using the leptonic CKM matrix (5) (or (A1) from Appendix) as
follows:
\be
\left\{ \ba{lc}
\nu_e(ct)=c_{13}\nu_1(0)e^{-i\varepsilon_1t}+
s_{12}c_{13} \nu_2(0)e^{-i\varepsilon_2 t}+
s_{13}\nu_3(0)e^{-i\varepsilon_3 t-i\delta'}\\
\nu_{\mu}(ct)=-(s_{12}c_{23}+s_{13}s_{23} e^{i\delta'})\nu_1(0)
e^{-i\varepsilon_1 t}+
c_{23}\nu_2(0)e^{-i\varepsilon_2 t}+c_{13}s_{23}\nu_3(0)e^{-i\varepsilon_3 t}\\
\nu_{\tau}(ct)=(s_{12}s_{23}-s_{13}c_{23} e^{i\delta'})\nu_1(0)
e^{-i\varepsilon_1 t}-
s_{23}\nu_2(0)e^{-i\varepsilon_2 t}+c_{13}c_{23}\nu_3(0)e^{-i\varepsilon_3 t}
\ea \right.
\ee
up to the terms of the second order in small quantities $s_{12},s_{13}\ll 1$.
(see Appendix for the exact $\hat V^l_{CKM}$ matrix) Here $t=L/c$ is the time
when neutrinos are observed at a distance $L=ct$ from their source; the
probabilities of neutrino observation at this distance from the
source is $P_{ab}(t)=|(\nu_a(t) \bar \nu_b(0))|^2$.
Multiplying $\nu_a(t)$ from Eq.(35) by $\bar \nu_b(0)$ and taking into account
the orthogonality of $\nu_a(0)$ states $(\nu_a(0)\bar \nu_b(0))=\delta_{ab}$,
it is easy to find:
\be
\left\{ \ba{rrl}
P(\nu_e\nu_e)&=&|c^2_{13}+s^2_{12}c^2_{13}e^{i\varphi_{21}}+
s^2_{13}e^{i\varphi_{31}}|^2\\
P(\nu_\mu \nu_\mu)&=&||s_{13}c_{23}+s_{13}s_{23}e^{i\delta'}|^2+
c^2_{23}e^{i\varphi_{21}}+c^2_{13}s^2_{23}e^{i\varphi_{31}}|^2\\
P(\nu_\tau\nu_\tau)&=&||s_{12}s_{23}-s_{13}c_{23}e^{i\delta'}|^2+
s^2_{23}e^{i\varphi_{21}}+c^2_{13}c^2_{23}e^{i\varphi_{31}}|^2\\
\ea \right.
\ee
\be
\left\{ \ba{rrl}
P(\nu_e\nu_\mu)&=&|c_{13}(c_{23}s_{12}+s_{23}s_{13}e^{i\delta'})
-s_{12}c_{13}c_{23}e^{i\varphi_{21}}-
s_{13}c_{13}s_{23}e^{i(\delta'+\varphi_{31})}|^2\\
P(\nu_e\nu_\tau)&=&|c_{13}(s_{23}s_{12}-c_{23}s_{13}e^{i\delta'})
-s_{12}c_{13}s_{23}e^{-i\varphi_{21}}+
s_{13}c_{13}c_{23}e^{i(\delta'+\varphi_{31})}|^2\\
P(\nu_\mu\nu_\tau)&=&|(s_{12}s_{23}-s_{13}c_{12}c_{23}e^{-i\delta'})
(s_{12}c_{23}+s_{13}c_{12}s_{23}e^{i\delta'})+
c_{23}s_{23}e^{i\varphi_{21}}\\&&
\hspace{8.2cm}-s_{23}c_{23}c^2_{13}e^{i\varphi_{31}}|^2
\ea \right.
\ee
where
\be
\varphi_{ij}=(\ve_i-\ve_j)L/c=\frac{\Delta m^2_{ij}}{2p_{\nu}}L=
2.54 \frac{L(m)}{E_\nu(MeV)}\Delta m^2(eV^2)
\ee
and
$$
\Delta m^2_{ij}=m^2_i-m^2_j,\quad i,j=1,2,3\quad
(\mbox{as }\ve_i \simeq p_{\nu}+\frac{m^2_i}{2p_{\nu}}\mbox{ at } cp\gg m_i).
$$
Eqs.(36)--(38) determine the neutrino oscillation probabilities in
vacuum ignoring the very important in some cases MSW effect
of the medium influence (see in ref.[10]). This effect has been
well studied and can be included separately (e.g. for
$\nu_e\nu_e,\:\nu_e\nu_\mu,\:\nu_e\nu_\tau$
in the solar neutrino case). The expressions for $P(\nu_{\mu}\nu_e),\:
P(\nu_{\tau}\nu_e),\: P(\nu_{\tau}\nu_{\mu})$ coincide with Eqs.(37) with
the substitution $\delta'\to -\delta'$

The simple  algebra allows one to reduce Eqs.(36),(37) to the partially known
[17 - 20] expressions containing the leptonic CP-phase $\delta'$ and given
below in Appendix.

Let us rewrite the Eqs.(36),(37) in much more simple forms
calculating numerically the coefficient in front of
$\cos \vp_{ij}=1-2\sin^2(\vp_{ij}/2)$
using for example the central values
of leptonic mixing angles
$\vt^l_{12},\vt^l_{13},\vt^l_{23}$ given in Eqs.(22), i.e. their sines
$s_{12},s_{13},s_{23}$, respectively. These equations in their numerical form
show clearly the influence of the leptonic CP-phase $\delta'$ on the neutrino
oscillations patterns:
\be
\left\{
\ba{r}
1-P(\nu_e\nu_e)\simeq
c_{12}^2 \sin^2(2\vartheta_{13}^l)\sin^2(\vp_{31}/2)+
c^4_{13}\sin^2(2\vartheta_{12}^l)\sin^2(\vp_{21}/2)\hspace{2.6cm}\\
+s_{12}^2\sin^2(2\vartheta_{13}^l)\sin^2(\vp_{32}/2)\simeq\\
0.23\sin^2(\vp_{31}/2)+4.3\cdot 10^{-3}\sin^2(\vp_{21}/2)
+0.29\cdot 10^{-3}\sin^2(\vp_{32}/2)\\
1-P(\nu_\mu\nu_\mu)\simeq
(0.94+0.017\cos\delta')\sin^2(\vp_{32}/2)
+(0.062-0.017\cos\delta')\sin^2(\vp_{31}/2)\\
(0.063+0.016\cos\delta'-0.31\cdot 10^{-3}\cos^2\delta')\sin^2(\vp_{21}/2)\\
1-P(\nu_\tau\nu_\tau)\simeq
(0.94+0.017\cos\delta')\sin^2(\vp_{32}/2)
+(0.062-0.017\cos\delta')\sin^2(\vp_{31}/2)\\
(0.064-0.016\cos\delta'-0.31\cdot 10^{-3}\cos^2\delta')\sin^2(\vp_{21}/2)\\
\ea \right.
\ee
Eqs.(37) determine the probabilities for neutrino of a given sort $\nu_i$ to
change its type (i.e. to transform into the other sort $\nu_j$) at a
distance L from the source of neutrinos. For $\nu_e$ this probability
does not depend on the CP-phase $\delta'$ at all as can be seen
already from Eqs.(36),(37), however, for transformations of
$\nu_{\mu}$ and $\nu_{\tau}$ this dependence is not too small.

The probabilities for different neutrino transitions
$P(\nu_i\nu_j)\ne P(\nu_j\nu_i)$ depend also on the value of $\delta'$:
\be
\left\{
\ba{r}
P(\nu_e\nu_{\mu})\simeq
0.059\{1-0.14\cos\delta'-0.98[\cos(\vp_{31})+0.019\cos(\vp_{21})
+0.0012\cos(\vp_{32})]\\
+0.140[\cos(\vp_{32}+\delta')-\cos(\vp_{31}+\delta')-\cos(\vp_{21}-\delta')]\}\\
P(\nu_e\nu_{\tau})\simeq
0.061\{1-0.13\cos\delta'-0.98[\cos(\vp_{31})+0.018\cos(\vp_{21})
+0.0012\cos(\vp_{32})]\\
-0.135[\cos(\vp_{31}+\delta')+\cos(\vp_{32}+\delta')-\cos(\vp_{21}-\delta')]\}\\
P(\nu_\mu\nu_\tau)\simeq
0.47\{1-0.99[\cos(\vp_{32})-0.066\cos(\vp_{21})+0.061\cos(\vp_{31})]
\hspace{2.6cm}\\
+0.035\sin\delta' \cos( \frac{\ds \vp_{31}+\vp_{32}}{\ds 2})
+0.035\sin\delta'\cos(\frac{\ds \vp_{31}+\vp_{32}}{\ds2})\sin(\vp_{21}/2)\\
-0.017\sin\delta'\sin(\vp_{21})\}\\
\ea
\right.
\ee

It is interesting to consider the time reversal effect which reveals in
the difference between the probabilities
$P(\nu_e\nu_{\mu})$ and $P(\nu_{\mu}\nu_e)$
or $P(\nu_e \nu_{\tau})$ and $P(\nu_{\tau} \nu_e)$ etc. [17].
Taking the difference between $P(\nu_e\nu_{\mu})-P(\nu_{\mu}\nu_e)$ and the same for
$\nu_e\nu_{\tau}$, $\nu_{\mu}\nu_{\tau}$ one obtains:
\be
\left\{
\ba{l}
P(\nu_{\mu}\nu_e)-P(\nu_e\nu_{\mu})=
0.0164(\sin\vp_{21}+\sin\vp_{32}-\sin\vp_{31})\sin\delta'\\
P(\nu_{\tau}\nu_e)-P(\nu_e\nu_{\tau})=
-0.0164(\sin\vp_{21}+\sin\vp_{32}-\sin\vp_{31})\sin\delta'\\
P(\nu_{\tau}\nu_{\mu})-P(\nu_{\mu}\nu_{\tau})=
0.0328(\cos\frac{\ds\vp_{21}}{\ds 2}-\cos\frac{\ds (\vp_{31}+\vp_{32})}{\ds 2})
\sin\delta'
\ea \right.
\ee
where $0.0328=c_{13}\sin 2\vt^l_{12}\sin 2\vt^l_{13}\sin 2\vt^l_{23}$
for the numerical coefficients given above. So, the leptonic
CP-phase can manifest itself in time reversal neutrino transitions
$\nu_a \nu_b$ and $\nu_b \nu_a$ experiments.
However, these experiments need the neutrino beams with a fixed value of
$L/2E_{\nu}$ what is very difficult to organize because usually one
deals with continuum spectra of produced neutrino.

We emphasize once more that the numerical coefficients in
Eqs.(39),(41) can be determined by future experimental data only,
but the general form of them (obtained algebraically from exact Eqs.(A2)--(A6)
of Appendix) remains always valid.

Let us average Eqs.(39),(40) over $L/2E_{\nu}$ considering the case of large
$L/2E_{\nu}>(\Delta m^2_{21})^{-1}$, i.e.  of large angles $\vp_{ij}=\Delta
m^2_{ij}L/2E_{\nu}$. This nicely corresponds to the real experimental situation
where one works with a continuum spectra of neutrino. This averaging gives:
\be
\ba{c}
\langle1-P(\nu_e\nu_e)\rangle\simeq 0.1193,\\
\langle1-P(\nu_\mu\nu_\mu)\rangle\simeq 0.529(1+0.016\cos\delta'-0.289\cdot
10^{-3}\cos^2\delta'),\\
\langle1-P(\nu_\tau\nu_\tau)\rangle\simeq 0.531(1-0.015\cos\delta'-0.288\cdot
10^{-3}\cos^2\delta')
\ea
\ee
and
\be
\ba{c}
\langle P(\nu_e\nu_\mu)\rangle\simeq 0.0585(1+0.140\cos\delta'),
\quad \langle P(\nu_e\nu_\tau)\rangle\simeq 0.0608(1-0.134\cos\delta'),\\
\langle P(\nu_\mu\nu_\tau)\rangle\simeq 0.470(1+0.394\cdot 10^{-3}\cos\delta'
-0.163\cdot 10^{-3}\cos2\delta')
\ea
\ee
where brackets mean the averaging over all $\vp_{ij}$.

Comparing Eqs.(39) and~(40) (and also (A2)--(A6) in the Appendix) one finds:
$P(\nu_i\nu_i)+P(\nu_i\nu_j)+P(\nu_i\nu_k)=1$ for different
$\nu_i,\nu_j$ and $\nu_k$ (e.g.
$ 1-P(\nu_e\nu_e) = P(\nu_e\nu_{\mu})+P(\nu_e\nu_{\tau}) $).

Eqs.(42)--(43) show also that the leptonic CP-phase $\delta'$ can be observed
experimentally, in principle, by measuring the average $\nu_e\nu_{\mu}$, or
$\nu_e\nu_{\tau}$ transition rates with 14\% accuracy which is much larger
than the effect (41) of time reversal.

This is illustrated in Figs. 2(a,b) where the averaged probabilities of
$\nu_i\nu_j$ transitions \mbox{$\langle P(\nu_i\nu_j)\rangle$} are calculated
for the values of $s_{12},s_{13},s_{23}$ determined in Eqs.(22). These
figures show a large (about 30\%) difference between $\langle
P(\nu_{\mu}\nu_e)\rangle$ and $\langle P(\nu_{\tau}\nu_e)\rangle$ probabilities
dependencies on the CP-phase $\delta'$. For example the value of
$\langle P(\nu_{\mu}\nu_e)\rangle\simeq 0.067$ at $\delta'=0$ turns out to be
larger than $\langle P(\nu_{\tau}\nu_e)\rangle\simeq 0.057$ by 17\%, i.e.
\mbox{$\frac{\ds \langle P(\nu_{\mu}\nu_e)\rangle}
{\ds \langle P(\nu_{\tau}\nu_e)\rangle}\simeq 1.17$},
while at $\delta'=\pi$ vice versa
$\langle P(\nu_{\mu}\nu_e)\rangle\simeq 0.050$ becomes smaller than $\langle
P(\nu_{\tau}\nu_e)\rangle\simeq 0.059$ by 15\% :  $\frac{\ds \langle
P(\nu_{\mu}\nu_e)\rangle}{\ds \langle P(\nu_{\tau}\nu_e)\rangle}\simeq 0.85$.

Unfortunately, the absolute values of these probabilities are small of about
$1/20$, but nevertheless they be really observed experimentally. Even weaker
is the $\delta'$ dependence of the average value of the probability of
$\nu_{\mu}\nu_{\tau}$ transition rate (averaged over the oscillations connected
with different $L/E_{\nu}$ values, or over $\vp_{ij}$ -- as in Figs. 2(a,b)),
shown in Fig. 2c. This figure shows that $\langle P(\nu_{\mu}\nu_{\tau})\rangle$
changes by 0.05\% only when $\delta'$ changes from $0$ to $\pi$. \\ \\

{\bf 5.}{\bf Conclusion}\\

The following three simple problems were discussed and solved in this paper:\\
a) The u- and d-quarks and electrons mass matrices were obtained in a simple
hierarchical form, including quark's CP-phase $\delta$ and correcting the
matrices suggested by B.Stech,\\
b) The recent data on neutrino masses and mixing angles were discussed shortly
and used for construction of the neutrino and leptonic mass matrices and CKM
matrix, both including the leptonic CP-phase $\delta'$,\\
c) The CKM matrix obtained was used to investigate the three neutrino
oscillations in the vacuum. The method of determination of the leptonic CP-phase
from their observation (averaged over the oscillations) was presented.

Note, that there is a vast ambiguity in the determination of quark's (or
leptonic) mass matrices. E.g. the pair of matrices $\hat M'_u=\hat U_o\hat
M_d\hat U_o^+$, with any unitary matrix $\hat U_o$, leads exactly to the same
mass eigenvalues and to the same mixing angles as $\hat M_u,\;\hat M_d$.

Also the following correction has to be added to the central part of the paper.
It was noted there (in Section 2, after Eq.~(21)) that Stech's matrices~(2),(3)
have negative value of some masses and that they positivity can be restored by
a simple $\gamma_5$ transformation. However this will violate the symmetry of
different generations fields and also will change the form of mass matrices.
Better is to avoid this shortcoming and construct the particles mass matrices
with only positive eigenvalues
$
\hat M^{diag}_a= \left(
\raisebox{9pt}{\makebox[18pt][c]{$m_{a_1}$}}
\raisebox{0pt}{\makebox[18pt][c]{$m_{a_2}$}}
\raisebox{-9pt}{\makebox[18pt][c]{$m_{a_3}$}}
\right)
$
directly from Eqs.~(11) [or~(29) and~(12)], e.g. for
d-quark and electrons one finds:
$$
\hat M'_d=
\left( \ba {rrl}a&\rho&0\\\rho&b&0\\0&0&m_b \ea\right),
\mbox{ where }
\left\{
\ba{l}
a=s^2_{12}m_2+c^2_{12}m_1
\simeq 15\sigma^3 m_b\\
b=c^2_{12}m_2+s^2_{12}m_1
\simeq c_{12}^2m_2
\simeq\frac{\ds 1}{\ds 2}\sigma^2m_b\\
\rho=-c_{12}s_{12}(m_2-m_1)
\simeq\sqrt{\frac{\ds 7}{\ds 2}}\sigma^2m_b,\mbox{~~~~or: }
\ea
\right.
$$
$$
\hat M_d'=
\left(
\ba{ccc}
15\sigma^4 & -\sqrt{\frac{\ds 7}{\ds 2}}\sigma^3 & 0\\
-\sqrt{\frac{\ds 7}{\ds 2}}\sigma^3 & \frac{\ds 1}{\ds 2}\sigma^2 & 0\\
0 & 0 & \sigma \ea\right)\frac{\ds m_b}{\ds \sigma}, \mbox{ ~~and similarly:~~ }
\hat M_e'= \left(\ba{ccc} 3\lambda'_1\sigma^4&\sqrt{\frac{\ds 3}{\ds 2}}\sigma^3&0\\
\sqrt{\frac{\ds 3}{\ds 2}}\sigma^3&(1+\beta')\sigma^2&0\\
0&0& \sigma\ea \right)\frac{\ds m_{\tau}}{\ds\sigma},
$$
with small $\lambda'=0.020,\;\beta'=0.016$. Both $\hat M_d,\;\hat M_e$ differs
slightly from Stech's forms given in Section 1.  The u-quark mass matrix is
determined by Eqs.~(12),(19) and also slightly deviates from the Stech's form
given by Eq.~(2):
$$
\hat M'_u=\hat U_u^+\hat M_u^{diag}\hat U_u\simeq
\left( \ba{ccc}
2\sigma^4 &\frac{\ds 1}{\sqrt{\ds 2}}\sigma^3\eta^+ &\sigma^2\eta^+ \\
\frac{\ds 1}{\sqrt{\ds 2}}\ds\sigma^3\eta & \frac{\ds 3}{\ds 2}\sigma^2 &
\frac{\ds 1}{\sqrt{\ds 2}}\sigma \\
\sigma^2\eta &\frac{\ds 1}{\sqrt{\ds 2}}\sigma & 1
\ea\right)m_t,\;
\mbox{ where }\hat U_u^+=\hat O_{23}^+\hat O_{13}^+(\delta)(\hat O_{12}^u)^+.
$$
The diagonalization of all these matrices
$\hat M_a^{diag}=\hat U_u\hat M_a'\hat U_u^+$ can be done by the same
unitary matrices $\hat U_d,\;\hat U_u,\;\hat U_e,\;\hat U_\nu$ with the same
mixing angles (used at they construction) as were used above for the Stech's
matrix case.

\newpage

\bc
{\bf Acknowledgments}\\
\ec

Authors express their gratitude to P.A. Kovalenko for calculation of u- and
d- running masses shown in Fig.1, to N. Mikheev for stimulating
discussion and also to L. Vassilevskaya and D. Kazakov for
useful discussions and an essential help in edition of the paper. Especially we
thank Z. Berezhiani who provided us with his and Anna Rossi paper [5] containing
a number of ideas used above. They thank the RFBR and INTAS for financial
supports by grants:  RFBR 96--15--96740, \mbox{98--02--17453}, INTAS 96i0155
and RFBR--INTAS:  96i0567, 95i1300. Also one of authors (K.A.\mbox{T-M})
acknowledges the Soros foundation for support by professor's grant.

\newpage

{\bf 6.} {\bf Appendix}\\
We give below the probabilities rates $P(\nu_i\nu_j)$ in general
algebraically form using the well known exact leptonic CKM matrix
(given above for quarks in Eq.(5)):\\
$$
\hat V^l_{CKM}= \left( \ba{ccl}
c_{12}c_{13}&s_{12}c_{13}&s_{13}e^{-i\delta'}\\
-s_{12}c_{13}-c_{12}s_{23}s_{13}e^{i\delta'}&c_{12}c_{23}-
s_{12}s_{23}s_{13}e^{i\delta'}&s_{23}c_{13}\\
s_{12}s_{23}-c_{12}c_{23}s_{13}e^{i\delta'}&-c_{12}s_{23}-
s_{12}c_{23}s_{13}e^{i\delta'}&c_{23}c_{13}\\
\ea
\right).
\eqno (A1)
$$\\
At $\delta'\ne 0$ it leads to:\\
$$
\ba{r}
1-P(\nu_\mu\nu_\mu)=
\{c^4_{23}\sin^2(2\vartheta^l_{21})+
s^4_{12}s^2_{13}\sin^2(2\vartheta_{23})+
s^4_{23}s^4_{13}\sin^2(2\vartheta^l_{12})+
c^4_{12}s^2_{13}\sin^2(2\vartheta^l_{23})\hspace{1.3cm}\\
+\cos\delta'\sin(4\vartheta^l_{12})\sin^2(2\vartheta^l_{23})
(s_{13}c_{23}^2-s^3_{13}s^2_{23})
-\cos^2\delta'
s^2_{13}\sin^2(2\vartheta^l_{23})\sin^2(2\vartheta^l_{12})\}
\sin^2(\vp_{21}/2)\\
+\{s^2_{12}c^2_{13}\sin^2(2\vartheta^l_{23})+
c^2_{12}s^4_{23}\sin^2(2\vartheta^l_{13})+
\cos\delta' s^2_{23}c_{13}
\sin(2\vartheta^l_{12})\sin(2\vartheta^l_{23})\sin(2\vartheta^l_{13})\}
\sin^2(\vp_{31}/2)\\
+\{c^2_{12}c^2_{13}\sin^2(2\vartheta^l_{23})+
s^2_{12}s^4_{23}\sin^2(2\vartheta^l_{13})-\cos\delta' s^2_{23}c_{13}
\sin(2\vartheta^l_{12})\sin(2\vartheta^l_{23})\sin(2\vartheta^l_{13})\}
\sin^2(\vp_{31}/2)
\ea
\eqno (A2)
$$
\vspace{0.5em}
$$
\ba{r}
1-P(\nu_\tau\nu_\tau)=
\{s^4_{23}\sin^2(2\vartheta^l_{21})+
s^4_{12}s^2_{13}\sin^2(2\vartheta^l_{23})+
c^4_{23}s^4_{13}\sin(2\vartheta^l_{12})+
c^4_{12}s^2_{13}\sin^2(2\vartheta^l_{23})\hspace{1.3cm}\\
+\cos\delta'\sin(4\vartheta^l_{12})\sin^2(2\vartheta^l_{23})
(s^3_{13}c^2_{23}-s_{13}s_{23}^2)-
\cos^2\delta's^2_{13}\sin^2(2\vartheta^l_{23})\sin^2(2\vartheta^l_{12})\}
\sin^2(\vp_{21}/2)\\
+\{s^2_{12}c^2_{13}\sin^2(2\vartheta^l_{23})+
c^2_{12}c^4_{23}\sin^2(2\vartheta^l_{13})-
\cos\delta'c^2_{23}c_{13}
\sin(2\vartheta^l_{12})\sin(2\vartheta^l_{23})\sin(2\vartheta^l_{13})\}
\sin^2(\vp_{31}/2)\\
+\{c^2_{12}c^2_{13}\sin^2(2\vartheta^l_{23})+
s^2_{12}c^4_{23}\sin^2(2\vartheta^l_{13})
+\cos\delta'c^2_{23}c_{13}
\sin(2\vartheta^l_{12})\sin(2\vartheta^l_{23})\sin(2\vartheta^l_{13})\}
\sin^2(\vp_{32}/2)
\ea
\eqno (A3)
$$\\
for $1-P(\nu_e\nu_e)$ see Eqs.(39). As before here
$\vp_{ij}=\Delta m^2_{ij}L/2E_{\nu}$ and:\\
$$
\ba{r} P(\nu_e\nu_{\mu})=\frac{\ds 1}{\ds 4}
\{\sin^2(2\vartheta^l_{13})(s^2_{23}+c^4_{12}s^2_{23}+s^4_{12}s^2_{23}+
\frac {\ds 1}{\ds 2}c_{13}\sin(2\vartheta^l_{23})\sin(4\vartheta^l_{12})
\cos\delta')\hspace{1.6cm}\\
-2c^2_{13}\sin^2(2\vartheta^l_{12})(c^2_{23}-s^2_{13}s^2_{23})\cos(\vp_{21})
-2s^2_{23}\sin^2(2\vartheta^l_{13})(c^2_{12}\cos(\vp_{31})+s^2_{12}
\cos(\vp_{32}))\\
+c_{13}\sin(2\vartheta^l_{12})\sin(2\vartheta^l_{13})\sin(2\vartheta^l_{23})
(s^2_{12}\cos(\delta'+\vp_{21})-c^2_{12}\cos(\delta'-\vp_{21}))\\
+c_{13}\sin(2\vartheta^l_{12})\sin(2\vartheta^l_{13})\sin(2\vartheta^l_{23})
(\cos(\delta'+\vp_{32})-\cos(\delta'-\vp_{31}))\\
+2c^2_{13}c^2_{23}\sin^2(2\vartheta^l_{12})\}
\ea
\eqno (A4)
$$
\vspace{0.5em}
$$
\ba{r}
P(\nu_e\nu_{\tau})=\frac{\ds 1}{\ds 4}
\{\sin^2(2\vartheta^l_{13})(c^2_{23}+c^4_{12}c^2_{23}+s^4_{12}c^2_{23}-
\frac {\ds 1}{\ds 2}c_{13}\sin(2\vartheta^l_{23})\sin(4\vartheta^l_{12})
\cos\delta')\hspace{1.6cm}\\
+2c^2_{13}\sin^2(2\vartheta^l_{12})(s^2_{23}-c^2_{13}s^2_{23})\cos(\vp_{21})
-2c^2_{23}\sin^2(2\vartheta^l_{13})(c^2_{12}\cos(\vp_{31})+s^2_{12}
\cos(\vp_{32}))\\
+c_{13}\sin(2\vartheta^l_{12})\sin(2\vartheta^l_{13})\sin(2\vartheta^l_{23})
(c^2_{12}\cos(\delta'-\vp_{21})-s^2_{12}\cos(\delta'+\vp_{21}))\\
+c_{13}\sin(2\vartheta^l_{12})\sin(2\vartheta^l_{13})\sin(2\vartheta^l_{23})
(\cos(\delta'+\vp_{31})-\cos(\delta'+\vp_{32}))\\
+2c^2_{13}s^2_{23}\sin^2(2\vartheta^l_{12})\}
\ea
\eqno (A5)
$$
$$
\ba{r}
P(\nu_{\mu}\nu_{\tau})=\frac{\ds 1}{\ds 4}
\{2s^2_{13}\sin^2(2\vartheta^l_{12})\cos^2(2\vartheta^l_{23})
+(c^4_{13}+c^4_{12}+s^4_{12}+(c^4_{12}+s^4_{12})s^4_{13})
\sin^2(2\vartheta^l_{23})\hspace{0.7cm}\\
-[2s^2_{13}(c^4_{23}+s^4_{23})\sin^2(2\vartheta^l_{12})+
[2s^2_{13}(c^4_{23}+s^4_{23})-(1+s^4_{13})\sin^2(2\vartheta^l_{12})]
\sin^2(2\vartheta^l_{23})]\cos(\vp_{21})\\
-[2c^2_{13}(s^2_{12}+c^2_{12}s^2_{13})\sin^2(2\vartheta^l_{23})-
\frac {\ds 1}{\ds 2}c_{13}\sin(2\vartheta^l_{12})\sin(2\vartheta^l_{13})
\sin(4\vartheta^l_{23})\cos\delta']\cos(\vp_{31})\\
+[2c^2_{13}\sin^2(2\vartheta^l_{23})(s^2_{12}s^2_{12}-c^2_{12})
-\frac {\ds 1}{\ds 2}c_{13}\sin(2\vartheta^l_{12})\sin(2\vartheta^l_{13})
\sin(4\vartheta^l_{23})\cos\delta']\cos(\vp_{32})\\
+2c_{13}\sin(2\vartheta^l_{12})\sin(2\vartheta^l_{13})
\sin(2\vartheta^l_{23})\sin\delta'\sin(\vp_{21}/2)
\cos(\frac{\ds\vp_{31}+\vp_{32}}{\ds 2})\\
+s_{13}\sin(4\vartheta^l_{12})\sin(4\vartheta^l_{23})\cos\delta'
[1+s^2_{13}]\sin^2(\vp_{21}/2)\\
-c_{13}\sin(2\vartheta^l_{12})\sin(2\vartheta^l_{13}\sin(2\vartheta^l_{23})
\sin\delta'\sin(\vp_{21})\\
-2s_{13}^2\sin(2\vartheta^l_{12})\sin(2\vartheta^l_{23})\cos(2\delta')
\sin^2(\vp_{21}/2)\}\\
\ea
\eqno (A6)
$$

For the average values of these probabilities over all $\vp_{ij}$ (i.e. over
$E_{\nu}$, or over $L$ at large $\vp_{ij}$) one obtains the values the (42),
(43) slightly dependent on the leptonic CP-phase $\delta'$:
$$
\left\{
\ba{rll}
\langle 1-P(\nu_e\nu_e)\rangle&=&A_{ee}\simeq 0.1193\\
\langle 1-P(\nu_{\mu}\nu_{\mu})\rangle&=&A_{\mu\mu}+B_{\mu\mu}\cos\delta'+
C_{\mu\mu}\cos^2\delta'\\
\langle 1-P(\nu_{\tau}\nu_{\tau})\rangle&=&A_{\tau\tau}+B_{\tau\tau}\cos\delta'+
C_{\tau\tau}\cos^2\delta'\\
\langle P(\nu_e\nu_{\mu})\rangle&=&A_{e\mu}+B_{e\mu}\cos\delta'\\
\langle P(\nu_e\nu_{\tau})\rangle&=&A_{e\tau}+B_{e\tau}\cos\delta'\\
\langle P(\nu_{\mu}\nu_{\tau})\rangle&=&A_{\mu\tau}+B_{\mu\tau}\cos\delta'
+C_{\mu\tau}\cos(2\delta')
\ea
\right.
\eqno (A7)
$$
where the coefficients are:
$$
\ba{rl}
A_{ee}=\frac{\ds 1}{\ds 2}
[c^2_{13}\sin^2(2\vartheta^l_{12})+\sin^2(2\vartheta^l_{13})]\hspace{0.95cm}
&\\
\ba{r}
A_{\mu\mu}=\frac{1}{2}
[(c^2_{13}+(c^4_{12}+s^4_{12})s^2_{13})\sin^2(2\vartheta^l_{23})\\
+(s^4_{13}\sin^2(2\vartheta^l_{12})+\sin^2(2\vartheta^l_{13}))s^4_{23}\\
+c^4_{23}\sin^2(2\vartheta^l_{12})]\\
\ea
&
\ba{r}
B_{\mu\mu}=-\frac{1}{2}s_{13}\sin(2\vartheta^l_{23})
(s^2_{23}s^2_{13}-c^2_{23})\sin(4\vartheta^l_{12})\\
C_{\mu\mu}=-\frac{1}{2}s_{13}\sin(2\vartheta^l_{23})
s_{13}\sin(2\vartheta^l_{23})\sin^2(2\vartheta^l_{12})
\ea\\
\ba{r}
A_{\tau\tau}=\frac{1}{2}
[(c^2_{13}+(c^4_{12}+s^4_{12})s^2_{13})\sin^2(2\vartheta^l_{23})\\
+(s^4_{13}\sin^2(2\vartheta^l_{12})+\sin^2(2\vartheta^l_{13}))c^4_{23}\\
+s^4_{23}\sin^2(2\vartheta^l_{12})]\\
\ea
&
\ba{r}
B_{\tau\tau}=-\frac{1}{2}s_{13}\sin(2\vartheta^l_{23})
(s^2_{23}-c^2_{23}s^2_{13})\sin(4\vartheta^l_{12})\\
C_{\tau\tau}=-\frac{1}{2}s_{13}\sin(2\vartheta^l_{23})
s_{13}\sin(2\vartheta^l_{23})\sin^2(2\vartheta^l_{12})
\ea\\
\ba{r}
A_{e\mu}=\frac{1}{4}
[(1+c^4_{12}+s^4_{12})s^2_{23}\sin^2(2\vartheta^l_{13})]\hspace{0.5cm}\\
+2c^2_{13}c^2_{23}\sin^2(2\vartheta^l_{12})]
\ea
&
\;\: B_{e\mu}=\frac{1}{8}c_{13}\sin(2\vartheta^l_{13})\sin(2\vartheta^l_{23})
\sin(4\vartheta^l_{12})\\
\ba{r}
A_{e\tau}=\frac{1}{4}
[(1+c^4_{12}+s^4_{12})c^2_{23}\sin^2(2\vartheta^l_{13})]\hspace{0.5cm}\\
+2c^2_{13}s^2_{23}\sin^2(2\vartheta^l_{12})]
\ea
&
\;\: B_{e\tau}=-\frac{1}{8}c_{13}\sin(2\vartheta^l_{13})
\sin(2\vartheta^l_{23})\sin(4\vartheta^l_{12})\\
\ba{r}
A_{\mu\tau}=\frac{1}{4}
[2s^2_{13}\sin^2(2\vartheta^l_{12})\cos^2(2\vartheta^l_{23})\hspace{1cm}\\
+\sin^2(2\vartheta^l_{23})\{(c^4_{12}+s^4_{12})s^2_{13}\\
+c^2_{13}+c^4_{12}+s^4_{12}\}]
\ea
&
\ba{l}
B_{\mu\tau}=\frac{1}{8}
(1+s^2_{13})s_{13}\sin(4\vartheta^l_{12})\sin(4\vartheta^l_{23})\\
C_{\mu\tau}=-\frac{1}{8}
s^2_{13}\sin^2(2\vartheta^l_{12})\sin^2(2\vartheta^l_{23})\\
\ea
\ea
\eqno (A8)
$$

\vspace{2ex}

\noindent The numerical  value of these coefficient are given above in the text
in (42), (43) for neutrino mixing angles $s_{12},\:s_{13},\:s_{23}$ values
determined in Eqs.(22).

\newpage

\bc
{\large \bf References.}
\ec
\begin{itemize}
\begin{description}
\item[\, 1.] See Particle Data Group, Eur.Phys.Journ. {\bf C3} (1998) 1.
\item[\, 2.] B.Stech\quad Phys.Lett. {\bf B403} (1997) 114. 
\item[\, 3.] S.Weinberg\quad Phys.Lett. (1979) 1566;\\
R.Barbieri, J.Ellis and M.K.Gaillard\quad Phys.Lett. {\bf B90} (1980) 249;\\
E.Akhmedov, Z.Berezhiani and G.Senjanovic,\quad Phys.Rev. Lett {\bf 69} (1992)
3013.
\item[\, 4.] For different see--saw types see K.  Ter--Martirosyan ITEP
preprint 93--88 (1988).
\item[\, 5.] Z. Berezhiani and A. Rossi, hep--ph/9811447\\
(the old H.Fritzsch approach was partly used in this paper see Nucl.Phys.
{\bf B155} (1979) 189).
\item[\, 6.] A.Ali Desy preprint 97--256, see  also hep--ph/9801270.
\item[\, 7.] J.A.K.Wermessen, S.A.Larin, T.van Ritberger\quad Phys.Lett.
{\bf B405} (1997) 327;\\
K.C. Chetyrkin\quad Nucl.Phys. {\bf B64}(1998)110
\item[\, 8.] P.A.Kovalenko\quad private communication.
\item[\, 9.] The Super Kamiokande Collab. hep--ex/9807003, hep--ex/9812009;\\
Y.Totsuka talk at the 18-th Internat. Symp. on Lepton and Photon Interaction,
 Hamburg, VII--VIII 1997;\\
Y.Fukuda et. al.,\quad Phys.Rev.Lett. {\bf 81} (1998) 1562, 2016.
\item[10.] N.Hata, P.Landgacker hep--ph/9705339;\\
J. Bahcall, P.Krastev, A.Smirnov hep--ph/9807216;\\
S.P.Mikheyev and A.Yu.Smirnov,\quad Yad.Fiz. {\bf 42} (1985) 1441;\\
L.Wolfenstein,\quad Phys.Rev. {\bf D17} (1985) 2369.
\item[11.] Y.Fukuda et. al.,\quad Phys.Lett. {\bf B.335} (1994) 237 (Kamiokande
data).
\item[12.] 
M.Apollonio et. al.(CHOOS Collab),\quad Phys.Lett. {\bf B420} (1998)
397;\\
M.Gonzalez--Garcia,\quad H.Nunokawa,\quad O.Peres and J.Valls hep--ph/9807305;\\
V.Berger, T.Weiler,  K.Whisnant hep--ph/9807319;\\
G.L.Fogli,  E.Lisi, A.Marone and G.Scioscia hep--ph/9808205;\\
S.Bilenky, C.Giunti and Grimus hep--ph/9809368.
\item[13.] R. Barbieri, L.Hall, A.Strumia, hep--ph/9808333; \\
R.Barbieri, L.Hall, D.Smith, A.Strumia, N.Weiner, hep--ph/9807235;\\
H.Fritzsch and Z.Xing, Phys.Rev. {\bf B440}(1998)313, hep--ph/9808272;\\
G.Altarelli and F.Feruglio hep--ph/9807353, hep--ph/9809596;\\
Z.Berezhiani, Z.Tavartkiladze,\quad Phys.Lett. {\bf B409} (1997) 220.
\item[14.\,] Z. Berezhiani and A. Rossi,\quad Phys.Lett. {\bf B367} (1996)
219;\\ C.D.Carone, L.Hall\quad Phys.Rev. {\bf D56} (1996) 4108;\\
M.Tanimoto\quad  Phys.Rev.  {\bf D57} (1998) 1983.
\item[15.\,] H.Georgi, S.L.Glashow preprint HUTP 98/0060 also hep--ph/9808233
v.2.
\item[16.\,] B.M.Pontecorvo JETP {\bf 33} (1957) 549, JETP {\bf 34} (1958)
247;\\
V.N.Gribov, B.M.Pontecorvo\quad Phys.Lett. {\bf 288} (1969) 483;\\
S.M.Bilenky, B.Pontecorvo\quad Phys.Rep. {\bf 41} (1078) 225.
\item[17\,] I.Yu.Kobzarev, B.V. Martemyanov, L.B.Okun, M.G.Shepkin\quad
Sov.J.Nucl.Phys.\\
{\bf 32}~(1980) 823, also ibid {\bf 35} (1982) 708.
\item[18.\,] V.Barger, Yuan-Ben Dai, K.Whisnant and Bing-Lin Yuong
hep--ph/9901388 v2, AMES-HET-99-01, ASITP-99-05 MADPH-99-1096, January-February
1999.
\item[19.\,] V.Barger, K.Whisnant and T.J.Weiler\quad  Phys.Lett. {\bf B427}
(1998) 97;\\
V.Barger, J.S.Pakvasa, T.J.Weiler and K.Whisnant,\quad Phys.Rev. {\bf D50}
(2998) 3016.
\item[20.\,] S.C.Gibbons, R.N.Mohapatra, S.Nandi and A.Raychauduri\quad
Phys.Lett.  {\bf B430} (1998) 296;\\
S.Mohanty, D.P.Roy and U.Sarkar, hep--ph/9810309.

\end {description}
\end{itemize}

\newpage

\bc
{\large \bf Figures captions.}
\ec
\begin{itemize}
\begin{description}
\item[Fig.1.] a) The run of the upper quark masses $m_t,m_c,m_u$
calculated in Refs.[8] in the first order of QCD perturbation theory (the
dashed lines) and in the fourth order of it (solid lines). The vertical line
corresponds to the scale $\mu =M_t\simeq 174.4$GeV used in the paper. The run
of electron's masses $m_{\tau},m_{\mu},m_e$ is disregarded (while it can
be easy taken into account and is not essential).\\
              b) The same for the masses $m_b,m_s,m_d$ of the lower
quarks.
\item[Fig.2.] a) Dependence of the average (over oscillations) value
$\langle P(\nu_e\nu_{\mu})\rangle$ of $\nu_e\nu_{\mu}$ transition probability on the
CP-phase $\delta'$; at $\delta'=\pi$ it has about 13\% minimum.\\
              b) The same for the average value $\langle P(\nu_e\nu_{\tau})\rangle$ of
$\nu_e\nu_{\tau}$ transition probability. As is seen instead of the minimum at
$\delta' =\pi$, as was for the $\nu_e\nu_{\mu}$ transition case, it has here
the 14\% maximum.\\
               c) The same dependence on the CP-phase as was shown in the
cases a), b) but for the average $\nu_{\mu}\nu_{\tau}$ probability
$\langle P(\nu_{\mu}\nu_{\tau})\rangle$. Its dependence on CP-phase $\delta'$ here is much
more flat -- about in hundred times smaller then in the $\nu_e\nu_{\mu}$ and
$\nu_e\nu_{\tau}$ cases.

\end {description}
\end{itemize}

\clearpage
\begin{figure}[t]
\bc
\bp(120,0)
\put(10,-20){\includegraphics{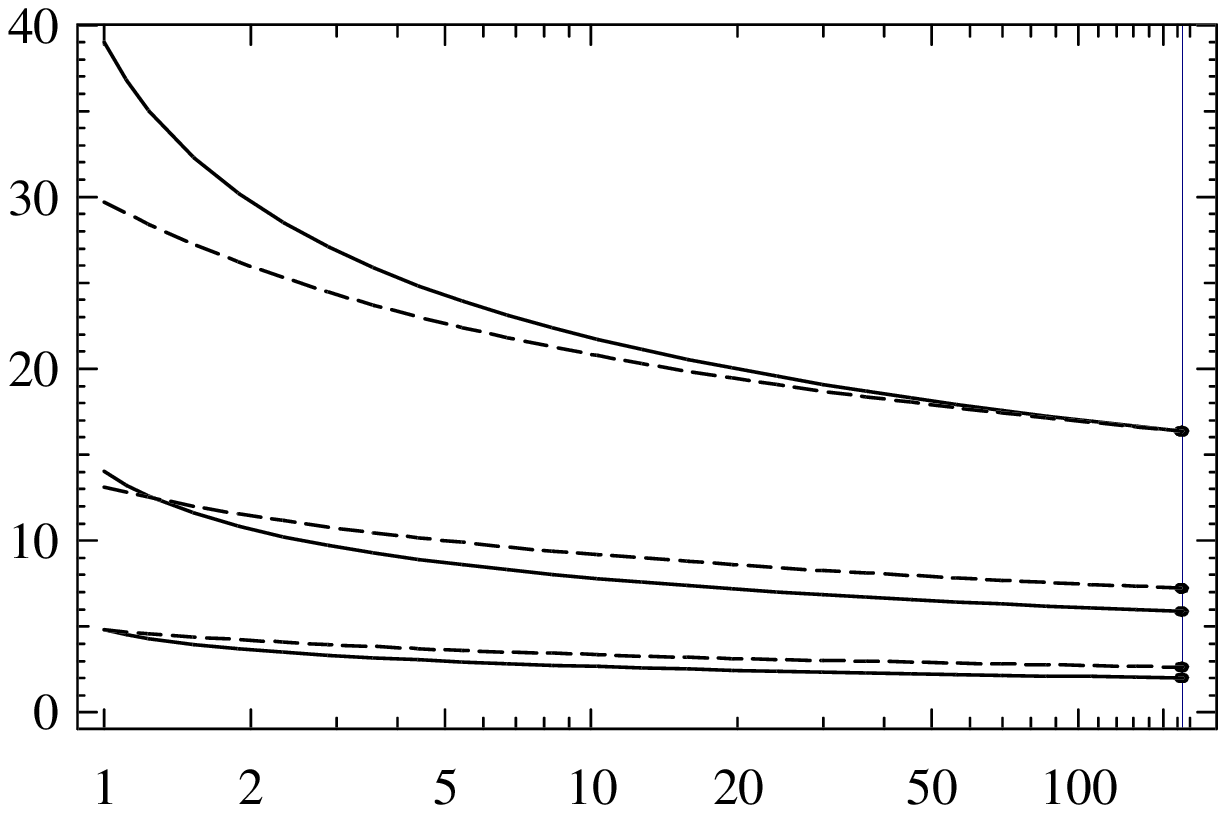}}
\put(107,-13.5){$\mu$,GeV}
\put(106,42.5){$m_t$}
\put(60,21){$m_t/10$}
\put(62,5){$10 m_c$}
\put(50,-2){$10^3m_u$}
\put(3,15){$GeV$}
\put(3,20){$m_i(\mu)$}
\put(115,15){$\large\bf a)$}
\ep
\ec
\end{figure}
\begin{figure}[b]
\bc
\bp(120,0)
\put(10,0){\includegraphics{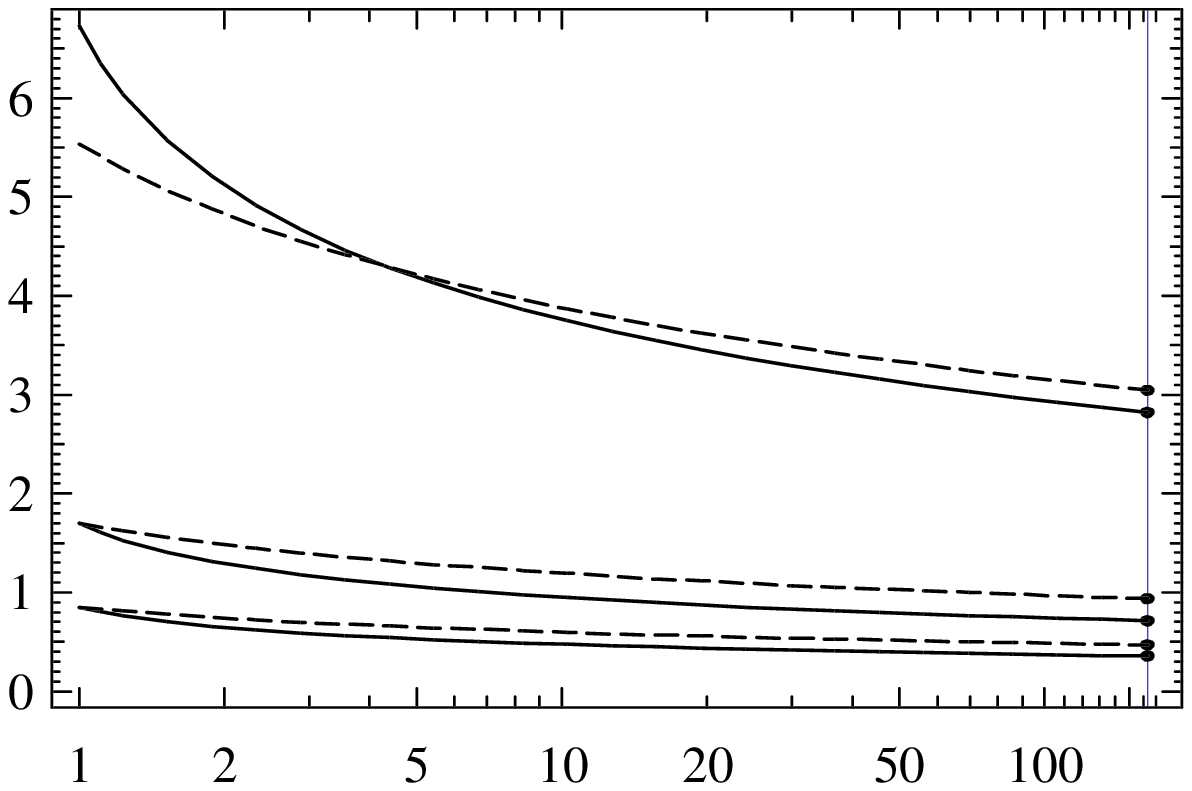}}
\put(107,-4.5){$\mu$,GeV}
\put(105,51.5){$m_t$}
\put(65,28.5){$m_b$}
\put(65,10.5){$10 m_s$}
\put(61,1.3){\small{$10^2m_d$}}
\put(115,26){$\large\bf b)$}
\put(3,26){$m_i(\mu)$}
\put(3,21){$GeV$}
\put(-20,56){Fig.1(a,b)}
\ep
\ec
\end{figure}

\begin{figure}[b]
\bc
\bp(120,0)
\put(-10,-33){\includegraphics{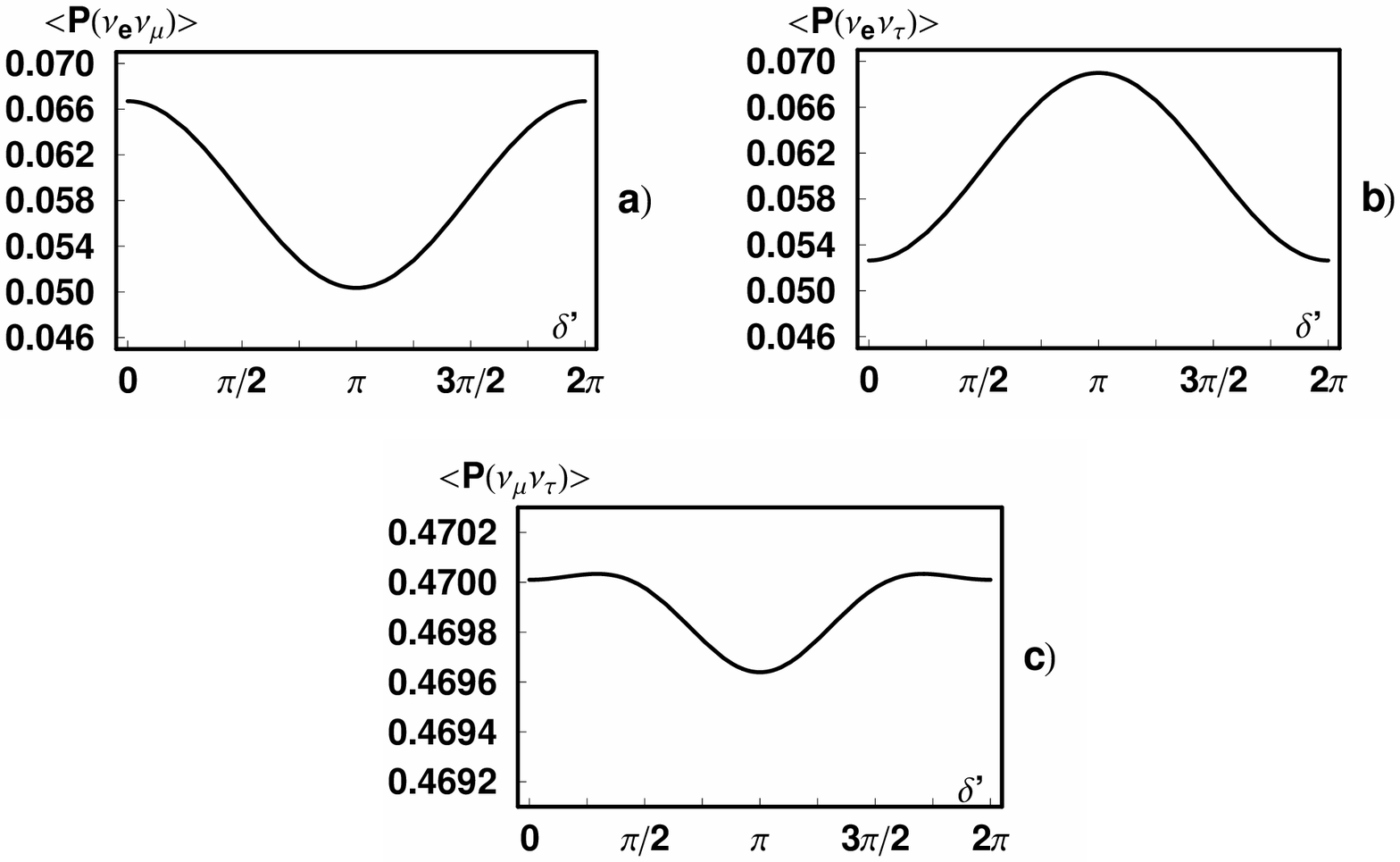}}
\put(-20,7){Fig.2(a,b,c)}
\ep
\ec
\end{figure}

\end{document}